 \documentclass[manuscript]{aastex}

\slugcomment{Submitted to the Astronomical Journal}
\shorttitle{Variable Stars in NGC~6316}
\shortauthors{Layden et al.}

\begin{document}


\title{Variable Stars in Metal-Rich Globular Clusters. II. NGC~6316}


\author{Andrew C. Layden\altaffilmark{1,2}}
\affil{Department of Physics \& Astronomy, 104 Overman Hall, 
Bowling Green State University, \\
Bowling Green, OH, 43403}
\email{layden@baade.bgsu.edu}

\author{Benjamin T. Bowes\altaffilmark{2}}
\affil{Department of Physics, RLM 5.208, University of Texas, Austin, TX, 78712}

\author{Douglas L. Welch\altaffilmark{1}}
\affil{Department of Physics \& Astronomy, McMaster University, \\
Hamilton, ON, L8S 4M1, Canada}

\and

\author{Tracy M. A. Webb\altaffilmark{1}}
\affil{Leiden Observatory, P.O. Box 9513, NL--2300 RA Leiden, The Netherlands}


\altaffiltext{1}{Visiting Astronomer, Cerro Tololo Inter-American Observatory.
CTIO is operated by AURA, Inc.\ under contract to the National Science
Foundation.}
\altaffiltext{2}{Previous address: Department of Astronomy, University
of Michigan, Ann Arbor, MI 48109}


\begin{abstract}

We present time-series $VI$ photometry of the metal-rich globular
cluster NGC~6316.  Our color-magnitude diagrams show a predominantly
red-clump horizontal branch morphology with no evidence of the blue
extensions seen in NGC~6388 and NGC~6441.  Our data are in good
agreement with published estimates of the reddening and metallicity
provided the cluster has an absolute distance modulus of 15.50 mag.  We
have discovered more than one dozen long-period variable stars in our field
of view, and argue that at least seven of them are members of
NGC~6316. We have also discovered four RR Lyrae variables, although only one
of them probably belongs to the cluster.  The specific frequency of RR
Lyrae stars in NGC~6316 is found to be less than 0.4, making it clear
that this cluster is a normal, metal-rich globular and is not an analog
of NGC~6388 and NGC~6441.  We recommend long-term photometric
monitoring of NGC~6316 to clarify the nature of the RR Lyrae member
candidate as well as the pulsation classes of the long-period
variables.
\end{abstract}



\section{Introduction}  \label{sec_intro}

Historically, the observational study of globular clusters has
provided critical tests of stellar structure and evolution theories.
In particular, the study of pulsating variable stars in globular
clusters has provided tests of stellar theory which complement the
matching of isochrones to observed cluster color-magnitude diagrams
(CMDs).  The types of pulsating variables found in globulars include
Long Period Variables (LPVs) on the asymptotic giant branch,
population II Cepheids, RR Lyrae variables (RRL) on the horizontal
branch, SX Phoenicis stars (population II $\delta$ Scuti stars) on the
main sequence or in the blue straggler region, and pulsating white
dwarfs.  Of these, the LPVs and RRL most easily accessible for study
because they are relatively luminous and are commonly found in
globular clusters.

Metal-rich globular clusters ([Fe/H] $> -0.8$), with their extremely
cool red giant branches (RGBs) and horizontal branches (HBs), are
expected to harbor large numbers of LPVs and few or no RRL.  NGC~104
(47 Tuc) is often considered to be the prototype.  However, two
important exceptions have been discovered. In NGC~6388 and NGC~6441,
\citet{rich97} observed blue HBs extending away from what appear to be
ordinary red HB clumps.  Subsequently, large numbers of RRL were
discovered in both of these clusters \citep{lrww99, pritzl01, pritzl02}.
The periods of these RRL were unusually long ($P > 0.7$ days),
suggesting that they are atypically luminous.

To date, both observers and theorists have been unable to identify a
definitive cause for the blue HB extensions, although many suggestions
have been put forward.  For example, \citet{rich97} suggested that age
or dynamical effects could be responsible for the blue HBs.  However,
no deep photometry has yet been published for these clusters to
constrain their ages, and calculations by \citet{rich97} indicated
that stellar interactions alone are too infrequent to explain the blue
HB extensions.  From the perspective of stellar theory, \citet{sc98}
have suggested that enhanced helium abundance and/or stellar rotation
could account for the blue HBs, but subsequent observational evidence
\citep{lrww99, msc99} has cast doubt on these explanations.  More
recently, \citet{rccbp02} and \citet{ryrl02} have suggested that the
stars in these clusters may have normal properties, but span a modest
range in age and/or metallicity, akin to the massive globular cluster
$\omega$ Centauri.  Photometry below the main sequence turn-off is
still required to test this hypothesis.  Clearly, the effort to
understand the cause of the blue HB extensions in NGC~6388 and
NGC~6441 is an ongoing and important endeavor.

A means of complementing this effort is to search for other globular
clusters which contain extended blue HBs.  Most of the metal-rich
globulars are located at low galactic latitudes, where differential
reddening and contamination by field stars in the disk and bulge
confuse the appearance of the CMDs, and can mask the presence of
extended blue HBs.  One way to alleviate this confusion is to search
for RRL -- old, pulsating HB stars -- based on their photometric
variability.  With this in mind, we have undertaken a variable star
survey of twelve of the metal-rich globulars listed by \citet{skk91}
as having poor or no variable star searches.  A fortunate byproduct of
this survey is the discovery of LPVs in these clusters.  \citet{fw98}
noted that the census of luminous LPVs (Mira variables) in globular
clusters is seriously incomplete.

In the first paper of this series, \citet{lrww99} discovered about 50
new variable stars toward NGC~6441, including at least eight RRL and
over 20 LPVs.  In this paper, we report our results for the second
cluster, NGC~6316 (C~1713--280).  According to \citet{harris96}, this
cluster is located at a heliocentric distance of 11.0 kpc in the
direction ($l$, $b$) = (357.18, +5.76) deg, behind the Galactic bulge.
The resulting foreground field star contamination and the high
reddening of $E(B-V) = 0.51$ mag present a challenge.  However, the
central concentration of the cluster (log $r_t / r_c$ = 1.55) is
moderate, so we can obtain reasonable photometry near the cluster
center.  Previous optical CCD photometry has been conducted by
\citet{arm88}, \citet{davidge92}, and \citet{hr99}, although the
cluster has never been surveyed for variable stars of any kind
\citep{clem01}.

In \S \ref{sec_obsreds} of this paper, we describe the observations and
data reduction procedures we employed, while \S \ref{sec_cmd} presents
our color-magnitude diagrams of the cluster and nearby field.  The
detection and lightcurve analysis of the variable stars is discussed
in \S \ref{sec_vars}.  In \S 5 we investigate the foreground reddening
of NGC~6316 in conjunction with its metallicity and distance.  We
consider whether the new variables are members of NGC~6316 in \S
\ref{sec_memb}, and present a summary of our results in \S \ref{sec_disc}.

\section{Observations and Reductions} \label{sec_obsreds}

We obtained time-series images of NGC~6316 using the direct CCD camera
on the 0.9-m telescope at Cerro Tololo Inter-American Observatory
(CTIO) during two runs in May and June of 1996 (3 and 8 usable nights,
respectively). The Tek\#3 2048 CCD provided a 13.5 arcmin field of
view with 0.4 arcsec pixels.  We used filters matched to the CCD to
reproduce the Johnson $V$ and Kron-Cousins $I$ bandpasses.
The raw images were processed by following the usual procedure for overscan
subtraction and bias correction, and by using twilight sky frames to
flat-field the images.
 
In a typical pointing toward the cluster, we obtained a short exposure
(40--120 sec) $VI$ frame pair and a long exposure (250--600 sec) $VI$
pair.  This provided two independent magnitude estimates of the HB
stars at each observational epoch, and extended the dynamical range of
the observations.  Such pointings were obtained 0--2 times each night,
with time intervals between pointings of at least two hours.  In
total, we obtained 47 images in twelve pointings toward NGC~6316.  The
seeing FWHM varied between 1.1 and 2.7 arcsec (1.7 arcsec median).  In
images taken during good seeing, a roughly radial gradient in the
quality of the stellar profiles was evident, and in some images, the
stars near the CCD corners were too far out of focus to yield reliable
photometry.


During a photometric night in 1996 May, we obtained a pointing toward
NGC~6316 along with an off-cluster $VI$ image pair centered 13.3
arcmin South of the cluster center.  The latter, which we will call
the control field, is outside the cluster tidal radius of 5.9 arcmin
\citep{harris96} and enables us to determine the photometric
properties of the foreground star field.  On this night, we also
obtained images of 63 \citet{aul92} photometric standard stars in 13
independent pointings.  The standards covered large ranges in color
($-0.3 < V-I < 4.0$) and airmass ($1.08 < X < 1.74$), and were
obtained at roughly hourly intervals throughout the night.  This time
($T$) coverage confirmed that the entire night was photometric, and
enabled us to seek and correct for small, slow variations in
extinction.  The instrumental aperture magnitudes of these stars
($A_f$), through filters $f=V$ or $I$, were fit with the equation
\[ A_f - S_f = c_{1,f} + c_{2,f} X + c_{3,f} (V-I) + c_{4,f} T + c_{5,f} T^2 
\]
where $S_f$ is the standard (Landolt) magnitude, and $c_{i,f}$
($i=1,5$) are the fitted coefficients.  The time-dependent terms
produced corrections of 0.01 mag or less.  The rms scatter about the
adopted fits were 0.010 and 0.013 mag in $V$ and $I$, respectively.

The cluster and control $VI$ pairs obtained on this night were reduced
using the {\sc Daophot} II point spread function (PSF) fitting
photometry package following the procedure outlined by \citet{pbs94}.
We selected $\sim$300 PSF stars well-distributed spatially over each
image to characterize the significant radial variation in PSF across
the image.  For each image, the PSF was iteratively improved by
fitting and removing faint neighbor stars.  This final PSF was applied
to the image using {\sc Allstar} to obtain optimal photometry for each
star on the image.  We then used {\sc Daophot} to subtract all the
fainter stars from each image, leaving only a set of several hundred
bright, isolated stars for use as secondary standards.  We measured
the aperture magnitudes of these stars using the same software and
parameters as was used for the Landolt standard stars, and used the
equations above to transform them to the \citet{aul92} $VI$ system.
These secondary standards were then used to transform the PSF-fit {\sc
Daophot} magnitudes of the fainter stars on the image onto the
\citet{aul92} scale.

The remaining images of NGC~6316, taken during non-photometric
conditions, were reduced using a version of the DoPHOT PSF-fitting
package \citep{sms93} which allows for spatial variations in the PSF.
Instrumental magnitudes were obtained for each image and combined into
instrumental $VI$ pairs, and then transformed to standard
\citet{aul92} $VI$ magnitudes via fits to the secondary standards
described above.  Each $VI$ image pair thus produced a single
photometry list.

The $XY$ positions from each list where aligned to a common coordinate
system,\footnote{A FITS-format image defining the ($X, Y$) system
used in Table \ref{tab_clusphot} is available at {\tt
http://physics.bgsu.edu/$\sim$layden/ASTRO/PUBL/published.html}.} and the
photometry of spatially coincident stellar images was combined using
an error-weighted mean.  As an estimate of the error in each combined
stellar magnitude, we adopted the standard error of the mean from the
individual measures.  For each star, we also computed the variability
index of \citet{ws93}, $I_{WS}$, from the individual magnitudes (see
\S \ref{sec_vars}).

Table \ref{tab_clusphot} presents the mean photometry for stars near
NGC~6316 that appear in at least four $V$ and four $I$ images, that
lie within eight arcmin of the cluster center, and that have a mean
magnitude of $V < 20.0$ mag.  Column 1 lists the identification
number, columns 2 and 3 are the $X$ and $Y$ pixel values of the star,
columns 4 and 5 are the mean $V$ magnitude and standar error of the
mean, columns 6 and 7 are the mean $I$ magnitudes and standard error,
columns 8 and 9 indicate the number of images on which the star
appeared, and column 10 lists the \citet{ws93} variability index,
$I_{WS}$.  Data for nearly 16,000 stars are listed.

The control field photometry consisted of only a single $VI$ image
pair with an intermediate exposure time of 300 sec.  These data are
presented in Table \ref{tab_fieldphot}; columns 1--7 are as in Table
\ref{tab_clusphot}.  Only stars within eight arcmin of the field
center, and with $V < 20.0$ mag are included.  Table \ref{tab_fieldphot}
contains data for over 15,000 stars.

\section{Color-Magnitude Diagrams}  \label{sec_cmd}

\begin{figure}
\plotone{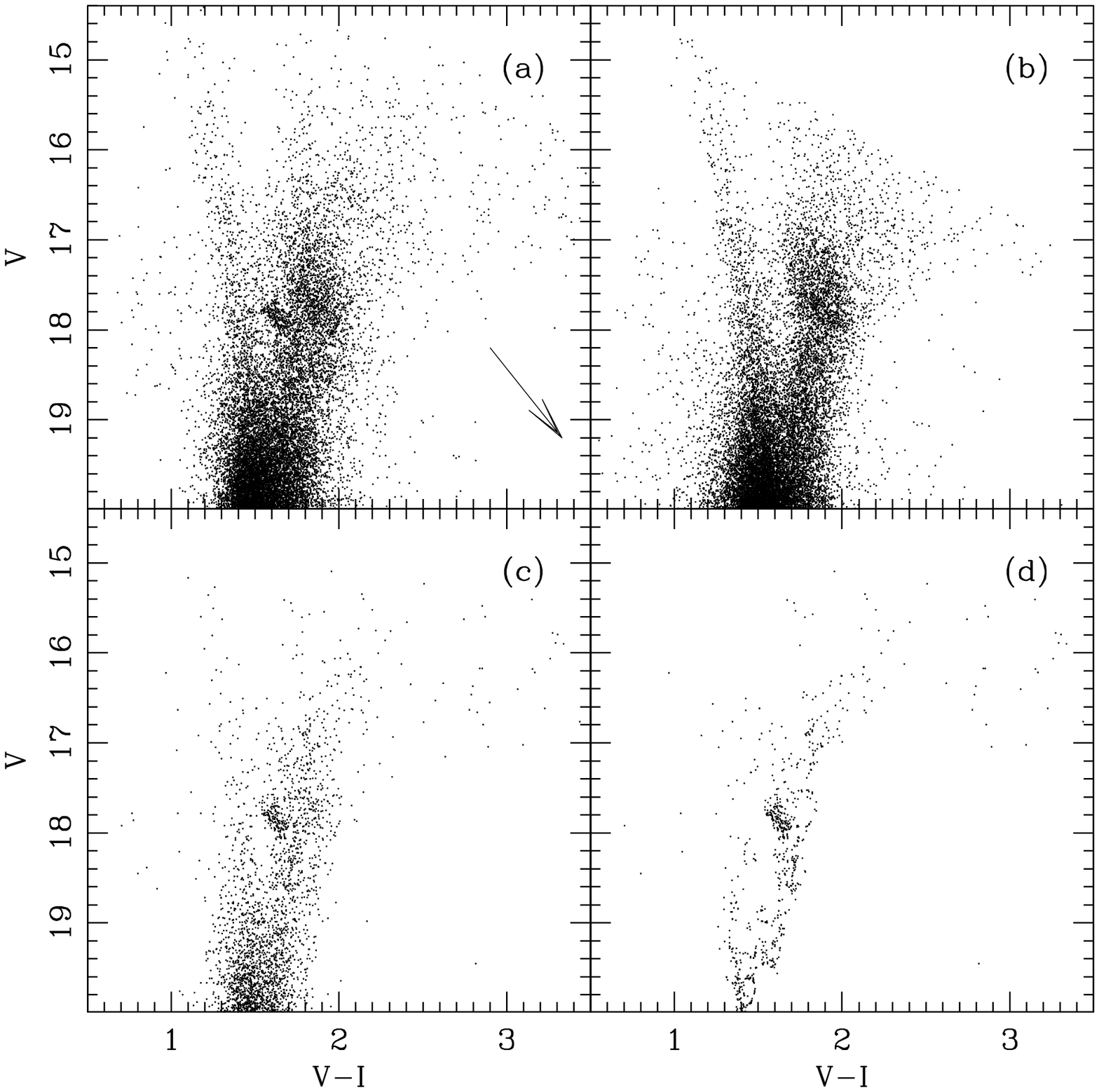}
\caption{Color-magnitude diagrams for stars with moderate errors
($\sigma_V < 0.10$ and $\sigma_{V-I} < 0.14$ mag) and located (a)
within eight arcmin of NGC~6316 [Table \ref{tab_clusphot}], (b) in the
control field [Table \ref{tab_fieldphot}], and (c) between 40 and 160
arcsec from the center of NGC~6316.  In (d), the points in (c) are
shown after the statistical subtraction of the field stars.  The arrow
indicates a reddening vector of $A_V = 1.0$ mag. \label{fig_cmd4}}
\end{figure}

The color-magnitude diagrams (CMDs) for the data in Tables
\ref{tab_clusphot} and \ref{tab_fieldphot} are shown in Figure
\ref{fig_cmd4}a and \ref{fig_cmd4}b, respectively.  Several features
are common to both panels, indicating that they represent the field
star populations (predominantly the Galactic disk and bulge) along the
line of sight through NGC~6316.  These include (a) the line of disk
main sequence stars extending from $V-I = 1.2$ and $V = 15.0$ mag to the lower
right, (b) the disk and bulge
red giant stars at $V-I > 1.7$ and $V < 19.0$ mag, (c) the bulge core
helium-burning stars (``red HB'' or ``red clump'') at $V-I \approx 1.9$
and $V \approx 17.6$ mag, and (d) the bulge blue HB stars at $V-I <
1.3$ and $17.0 < V < 18.8$ mag.  The only feature of the cluster CMD
that is clearly absent in the control field CMD is the group of stars
at $V-I = 1.65$ and $V = 17.9$ mag, which corresponds to the red HB
clump of NGC~6316.

The influence of the overwhelming field star population is reduced in
Figure \ref{fig_cmd4}c, where we plot only the stars located at a
projected radius between 40 and 160 arcsec from the center of
NGC~6316.  The cluster red clump is again evident, and appears
slightly elongated along the reddening vector, suggesting there is a
small amount of differential reddening across the face of the cluster.
This helps to explain why the cluster RGB remains ill-defined in this
panel.  The number of BHB stars is greatly reduced relative to Figure
Figure \ref{fig_cmd4}a, indicating that few, if any, are members of
NGC~6316. 

Another means of isolating the cluster stars from the field involves
the technique of statistical subtraction.  We generated a field star
CMD using only stars located between 40 and 160 arcsec of the center
of the control field image.  For each star in this CMD, we removed
from the cluster CMD (Figure \ref{fig_cmd4}c) the star nearest in
color-magnitude space.\footnote{Tests showed no detectable difference
in mean reddening between the cluster and control field CMDs.}
Statistically, the remaining stars should represent the locus of stars
in the cluster alone.  This CMD is shown in Figure \ref{fig_cmd4}d.
Though the subtraction is over-aggressive for the faint stars, the red
clump and lower RGB stars are more clearly shown.  Unfortunately,
subtraction was not possible for the upper RGB stars because they were
saturated in the control field images.  An important result of the
subtraction is that only a handful of blue HB stars remain, further
indicating that NGC~6316 does not have a well-populated blue HB like
the metal-rich globulars NGC~6388 and NGC~6441.  Specifically, we
would expect about 22 HB stars with $(V-I) < 1.55$ if NGC~6316 had the
same ratio of HB to red clump stars as seen in NGC~6441
\citep{lrww99}, whereas at most six can be seen in Figure
\ref{fig_cmd4}d.

\begin{figure}
\plotone{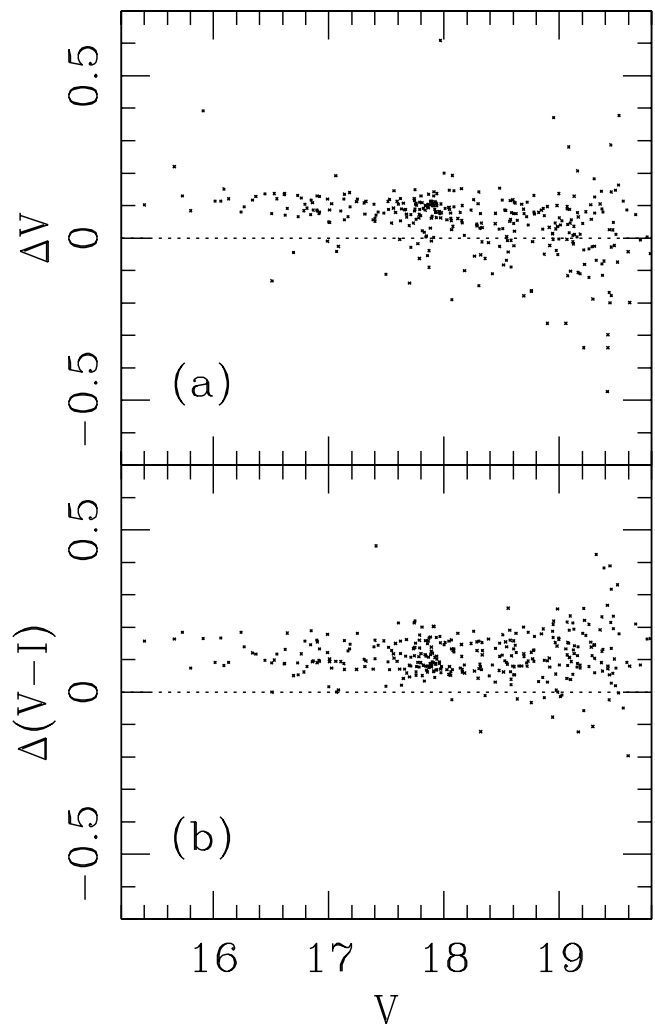}
\caption{A star-by-star comparison between our photometry and that
of \citet{arm88} in the sense ``ours - Armandroff's''.  (a) $V$
magnitude, and (b) $V-I$ color. \label{fig_armand88}}
\end{figure}

We can compare our photometry with other CCD photometry available in
the literature.  Figure \ref{fig_armand88} shows a star-by-star
comparison with the $VI$ photometry of \citet{arm88}.  Small but
significant systematic differences are apparent.  Considering bright
stars ($V < 18$ mag) and rejecting outliers, we find $\overline{\Delta
V} = 0.086 \pm 0.003$ mag (rms = 0.040 mag) and $\overline{\Delta
(V-I)} = 0.109 \pm 0.003$ mag (rms = 0.043 mag), with our photometry
being fainter and redder than that of \citet{arm88}.  The differences
do not appear to be color dependent, though there is a slight
correlation with $X$ coordinate.  

\citet{hr99} also presented $VI$ photometry of NGC~6316.  Estimating
from their observed CMD (their Figure 8), they find the typical color
and magnitude of the red clump to be 1.30 and 17.5 mag, respectively.
The median color and magnitude of the red clump in our data (Table
\ref{tab_fieldphot}) are 1.63 and 17.85 mag, while the values from
Armandroff's data were 1.56 and 17.80 mag.  \citet{davidge92} found
the red clump at $V \approx 17.85$ mag.

Clearly, there are disagreements between these studies, particularly
among the colors, and it is not obvious which photometric zeropoint
is correct.  Errors in aperture corrections due to the high stellar
density in the region are a likely cause of the discrepancies.  It is
worth noting that our larger areal coverage enables us to obtain
aperture corrections from stars farther from the crowded cluster
center.  For the remainder of this paper we adopt the data presented
in Table \ref{tab_clusphot}, but acknowledge the uncertainty in its
photometric zeropoint.

\section{Variable Stars}  \label{sec_vars}

We used the \citet{ws93} variability index, $I_{WS}$, to characterize
the likelihood that a given star varies in brightness.  The location
of points in the ($I_{WS}$, $V$) plane led us to classify stars as
probable variables ($I_{WS} > 130$) and possible variables ($30 <
I_{WS} < 130$).  When plotted on the cluster CMD, three classes of
variables were evident: extremely red variables (candidate long period
variables), blue variables (candidate RRL), and a few variables
scattered around the CMD (candidate eclipsing binaries or other
types).

The cluster center is rather crowded, resulting in larger photometric
errors and an increased likelihood of detection incompleteness for
faint or low-amplitude variables.  Variable star detection using
image-subtraction software \citep{alard00}, rather than PSF-fitting
software, may provide a more complete census of variables near the
cluster center.  In this approach, a number of high-quality images are
registered and co-added to produce a reference image.  This reference
image is convolved with a spatially-variable kernel to match the
seeing in each observed image, and the two images are subtracted.  The
non-variable stars subtract away, leaving positive or negative
residuals at the positions of variable stars.  We used the ISIS2.1
software by \citet{alard00} on our $V$-band images.  As described
below, the results confirmed the variability of stars found using
DoPHOT, and detected some additional candidate variables.



\subsection{Long Period Variable Stars}  \label{sec_lpv}

\begin{figure}
\plotone{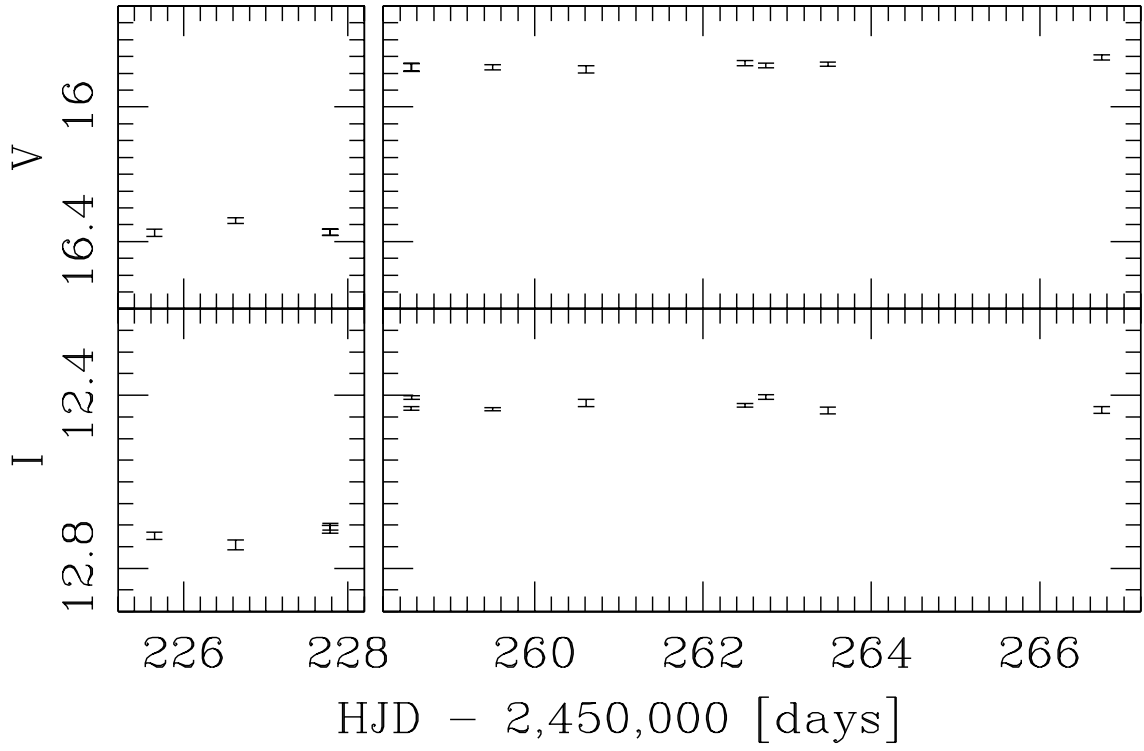}
\caption{Magnitude as a function of time for the long-period variable
star V4, a typical LPV in our study.  The panels on the left and right
represent the May and June (1996) observing runs, respectively.
\label{fig_lpv04}}
\end{figure}

When their $V$ and $I$ magnitudes were plotted as a function of time,
many of the red variable candidates showed constant or slowly-varying
apparent brightness, with distinct jumps between the May and June
data.  This suggests that they are long period variables (LPVs) of the
Mira, Semi-Regular, or Irregular classes.  An example of their
behavior is shown in Figure \ref{fig_lpv04}.  

Mean photometry for the thirteen detected LPV stars is presented in
Table \ref{tab_lpvphot}, where we have designated these stars V1
through V13.  This table also contains the identification number from
Table \ref{tab_clusphot} (``ID''),\footnote{The $XY$ coordinates of
each variable star in Tables \ref{tab_lpvphot}, \ref{tab_suspvar}, and
\ref{tab_rrlphot} can be found using the cross-reference ``ID'' number
listed in Table \ref{tab_clusphot}.  The $XY$ coordinates can be
converted to approximate right ascension and declination offsets from
the cluster center (in arc seconds, with positive values indicating
East and North) via the equations $\Delta \alpha \approx
-0.396~(X_{\rm pix}-1059)$, and $\Delta \delta \approx -0.396~(Y_{\rm
pix}-993)$.  \citet{harris96} lists the J2000.0 equatorial coordinates
of NGC~6316 as ($\alpha$, $\delta$) = (17:16:37.4, --28:08:24).}  the
radial distance of the variable from the cluster center in arcsec
($R$); the mean $I$ magnitude and $(V-I)$ color of the variable during
the May observing run ($\overline{I}_M$ and $\overline{V-I}_M$), and
the number of $VI$ observations obtained in May ($N_{M}$).  The next
two columns contain the difference between the average magnitudes and
colors determined in the May and June observing runs ($\Delta
\overline{I}_{M-J}$ and $\Delta \overline{V-I}_{M-J}$), and the number
of $VI$ observations obtained in June ($N_J$).

The ISIS2.1 analysis confirmed the variability of V1 through V13, and
suggested that several other stars might vary.  The DoPHOT
photometry of these stars was often compromised by light from blended
neighbor stars, resulting in time-magnitude plots with enhanced
scatter.  These stars are listed in Table \ref{tab_suspvar} as
SV1-SV6, along with their ``ID'' number and radial distance, $R$, as
in Table \ref{tab_lpvphot}.  Also listed are the mean $V$ magnitude
and $(V-I)$ color of the variable ($\overline{V}$ and
$\overline{V-I}$), the range of observed $V$ and $I$ magnitudes
($\delta V$ and $\delta I$), and the number of $VI$ pairs observed
($N_{obs}$).

The time-series photometry of the probable and suspected LPVs are
presented in Table \ref{tab_timeser}.  The columns contain: the
variable star name, the heliocentric Julian Date of the observation
(HJD), the observed $V$ magnitude and its error ($\sigma_V$), the
observed $I$ magnitude and its error ($\sigma_I$), the fitted light
curve phase ($\phi$) when applicable (see Sec. \ref{sec_rrl}), and a
quality code ($Q$).  This code is defined as: $Q = 1$ indicates a long
exposure under good seeing conditions ({\it filled circle}), $Q = 2$
indicates a short exposure in good seeing ({\it filled triangle}), $Q
= 3$ indicates a long exposure in poor seeing ({\it open circle}), and
$Q = 4$ indicates a short exposure in poor seeing ({\it open
triangle}).  Variability detected only in the ISIS2.1 analysis are
indicated by $Q = 5$.


Unfortunately, our observations do not sample the complete light
cycles of the LPVs, and thus do not accurately represent the stars'
mean photometric properties.  The limited phase coverage also leads to
LPV detection incompleteness, especially for stars which vary on 
long time scales or with low amplitudes.  The LPV stars in NGC~6316
deserve long-term photometric monitoring, though the detections
described here are an important first step in their study.

\subsection{RR Lyrae Variables and Eclipsing Binaries} \label{sec_rrl}

When plotted as a function of time, the magnitudes of the blue
variable candidates usually showed large variations on time scales
of a day or less.  The same was true for most of the variables
scattered around the CMD.  This type of behavior is expected from
RR Lyrae and close binary stars.  

The rapid variability makes it likely that, even with only twelve
independent observational epochs, we can obtain sufficient phase
coverage to determine the period and construct a light curve for a
candidate variable.  This information enables us to classify the
type of variability with some confidence.

We searched for periods using the template-fitting method described in
\citet{lrww99} and \citet{ls00}.  Briefly, this technique involves
folding a star's magnitude--time data by a sequence of periods, and at
each period, fitting the resulting light curve with a set of templates
describing the light curve shapes of different types of variables
(e.g., RRab, RRc, W UMa, etc.)  If the period being tested is near the
true period, the scatter about the fitted template will be small,
resulting in a small value of $\chi^2$.  Probable periods can thus be
identified by searching for $\chi^2$ minima in period--template
space.

We fit each star with ten templates, including curves for six RRab
with different shapes, one RRc, a contact binary, a detached binary,
and a cosine function \citep{ls00}.  We fit the templates over periods
ranging from 0.1 to 3.0 days (periods longer than 3 days are evident
in the time-magnitude plots).  To complement this analysis, we also
searched for periods in the range $0.2 < P < 2.0$ days using both the
Lomb-Scargle method \citep{sca82} and the String Length method (von
Braun \& Mateo, private communication).  In most cases, the periods
from the different methods agreed well.  

\begin{figure}
\plotone{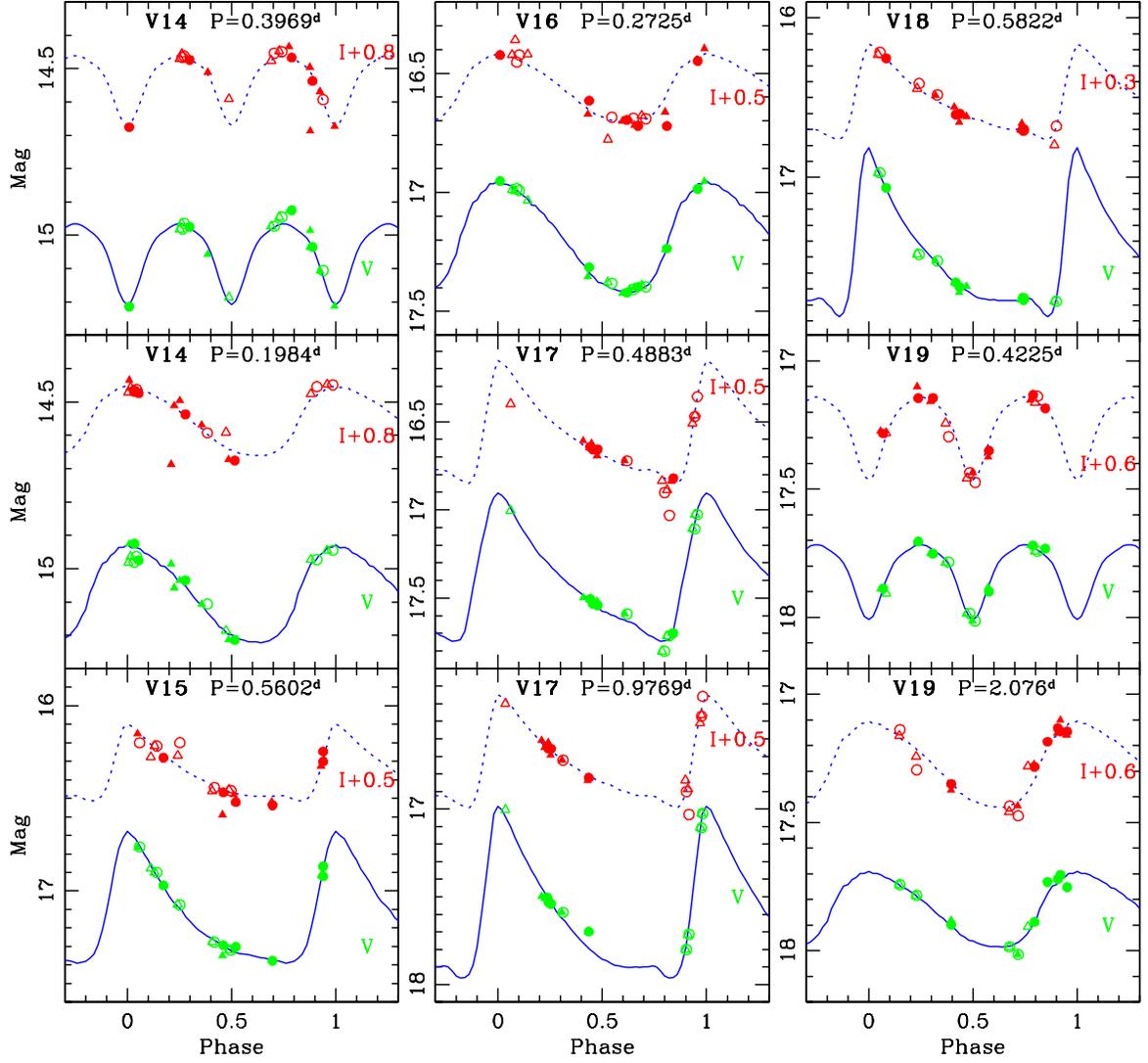}
\caption{Light curves for the short period variables.  Each panel
shows the $V$ and $I$ magnitudes of a specific variable folded by the
given period.  In some panels, the $I$-band data are shifted
vertically by the indicated amount for clarity.  The symbols indicate
the exposure length and seeing as described in \S 4.1.  The solid
and dashed curves are the best fitting template for the $V$ and
$I$-band data, respectively.  
\label{fig_9rrlc}}
\end{figure}

We determined periods for six of the short-period variable candidates.
Three appear to be RR Lyrae stars pulsating in the fundamental mode
(RRab), one is an RR Lyrae pulsating in the first overtone (RRc), and
two are probably contact binaries of the W Ursae Majoris type (WUMa).
Details of these stars' photometry are presented in Table
\ref{tab_rrlphot}, including the variable star designation, ID number
corresponding to Table \ref{tab_clusphot}, the radial distance from
the cluster center in arcseconds ($R$), the best-fitting period in
days ($P$), the class of variability corresponding to the best-fitting
template, the rms scatter of the observed points about that template
($rms_V$ and $rms_I$), the intensity-mean $V$ and $I$ magnitudes
($\langle V \rangle$ and $\langle I \rangle$), the light curve
amplitude ($\Delta V$ and $\Delta I$), the number of paired $VI$ data
points observed for the variable ($N_{obs}$), and a comment.  The
intensity means and amplitudes were determined from the best-fitting
template since the phase coverage of the actual observations was often
incomplete.  The light curves, along with the best-fitting templates,
are shown in Figure \ref{fig_9rrlc}.

For the stars V14, V17, and V19, the light curve fitting procedure
yielded two $\chi^2$ minima of comparable depth, and hence two
possible periods.  In the cases of V14 and V19, this led to ambiguity
in the variable class (contact binary versus pulsating star).  For
each of these stars, entries for both periods are given in Table
\ref{tab_rrlphot}, and light curves for both periods are shown in
Figure \ref{fig_9rrlc}.  The rms values listed in Table
\ref{tab_rrlphot} indicate the relative merit of the two periods.  The
star labeled SV2 in Table \ref{tab_suspvar} exhibited short period
variability in its magnitude--time plots, but did not present any
$\chi^2$ minima which yielded a satisfactory light curve.  Further
observations of these stars are required to determine their correct
periods and variable classes.  Time-series photometry for each of the
suspected and probable variable stars is given in Table
\ref{tab_timeser}.

We did not detect any detached eclipsing binaries in the region of
NGC~6316.  We suspect this is due largely to observational bias.  With
only twelve independent visits to the cluster, the chance of observing
a detached binary during eclipse is relatively small, and the chance
of observing enough eclipses to determine a unique period is smaller
still.  In Paper I, we found the detached eclipsing variables were not
members of the cluster, so their absence in this study of NGC~6316
should not adversely affect our inventory of variable stars in the
cluster.  We emphasize that our sampling is much more sensitive to
variables that exhibit continuous variations.  The ISIS2.1 analysis
confirmed the variability of the short-period variables V14--V19 and
did not detect any additional meaningful candidates near the crowded
cluster center.

\section{Interstellar Reddening}   \label{sec_redden}

We used the equations of \citet{card89} to establish the relations
between reddening and extinction in the filter passbands used in this
study.  We find $A_V/E(V-I) = 2.4$ and $E(V-I)/E(B-V) = 1.24$, where
$E(B-V)$ corresponds to the system defined in \citet{card89}.  The
reddening vector in Figure \ref{fig_cmd4} employs this relation.  The
cluster HB in that figure is elongated by differential reddening along
the same slope, indicating that $A_V/E(V-I) = 2.4$ is appropriate for
our data.

\citet{hr99} showed that there is a small amount of differential
reddening along the line of sight to NGC~6316.  They used the
cluster's red clump and RGB stars to derive a reddening map for stars
in a $2.24\times2.24$ arcmin box centered on the cluster (see their
Figure 22).  Having noted in \S \ref{sec_cmd} the difference in
photometric zeropoint between their data and ours, it seems that a
{\it differential} application of their map to our data is
appropriate.  We converted their map into a differential map by
subtracting the mean reddening in their field, 0.66 mag, from each map
value.\footnote{Experiments with our own data indicate a differential
reddening map with qualitatively similar behavior.}  

\citet{harris96} lists the metallicity and mean reddening of NGC~6316
as [Fe/H] = --0.55 and $E(B-V) = 0.51$ mag, respectively.  The
reddening value corresponds to $E(V-I) = 0.63$ mag.  Figure
\ref{fig_varcmd} shows the dereddened CMD of stars from Table
\ref{tab_clusphot} that are located within the $2.24\times2.24$ arcmin
box defined by \citet{hr99}.  The photometry of each star has been adjusted
using the differential reddening map derived from \citet{hr99}, and then
shifted by the mean reddening value from \citet{harris96}.

We can use isochrones in conjunction with our data to test whether the
\citet{harris96} reddening and metallicity values are correct.
\citet{bertelli94} provides isochrones that bracket the metallicity of
NGC~6316: $z = 0.004$ and 0.008 ([Fe/H] = --0.7 and --0.4 dex,
respectively).  Figure \ref{fig_varcmd} shows the 12 Gyr isochrones
($\log t = 10.08$) at these metallicities (isochrones with 10 and 14
Gyr ages are nearly identical).  They have been shifted vertically by
15.5 mag to match the red clump stars.  The isochrones bracket the
observed red clump and the upper RGB and AGB stars, supporting the
\citet{harris96} reddening and metallicity values.  The elongation of
the red clump suggests that differential reddening, on scales that are
unresolved by the \citet{hr99} map, continues to affect our data at
the level of $<$0.1 mag.  Alternatively, this elongation could
indicate a metallicity spread of $\sim$0.4 dex within the cluster.
However, the narrowness of the upper RGB and AGB sequences argues
against this.  Finally, we note that our photometry shows an
enhancement in the number of stars at $V_0 \approx 15.3$ mag, where
the isochrones double-back upon themselves along the AGB.

To summarize, our data are in excellent agreement with
globular cluster aged \citet{bertelli94} isochrones with the current
estimates of metallicity and reddening: [Fe/H] = --0.55 and $E(B-V) =
0.51$ mag.  The one free parameter in this analysis is the cluster
distance modulus, which we find to be $(V-M_V)_0 = 15.50 \pm 0.10$
mag.  This corresponds to a distance of $12.6 \pm 0.6$ kpc, which is
somewhat larger than the value of 11.0 kpc listed by \citet{harris96}.


\section{Cluster Membership}  \label{sec_memb}

The high density of field stars along the line of sight through
NGC~6316 makes it difficult to determine which variable stars are
members of the cluster and which are not.  Information on cluster
membership can be derived from the positions of the stars in the
CMD and the radial locations of the stars with respect to the
cluster center. Future radial velocity measurements are another
possible discriminant for membership.

The colors and magnitudes of all the variable stars are plotted on the
CMD shown in Figure \ref{fig_varcmd}.  The four LPVs within 3.0 arcmin
of the cluster center define a tight sequence, suggesting that most of
them are cluster members.  The LPV with $(V-I)_0 \approx 1.4$ (V12)
appears to be too faint and blue to be a pulsating AGB star, but could
be a contact eclipsing binary involving a red giant star.  The LPVs
V7, V11, and V12 are very close to the cluster's tidal radius of 5.9
arcmin, and so are unlikely to be members of NGC~6316.  Of the five
suspected LPVs, only SV1 appears capable of being a cluster member.
Membership of the stars redder than $(V-I)_0 = 3.2$ mag is uncertain
since the isochrones leave some ambiguity over the exact position and
extent of the AGB tip.  Based on this information, we classified as
LPV members (commented with an ``m'' in Table \ref{tab_lpvphot}) the
seven variables with $R < 3.0$ arcmin and which have $2.4 < (V-I)_0 <
3.2$ mag.  The group of LPVs at ($V-I$, $V$) = (4.0, 15.6), which
includes V8, V10, and SV1, are possible members.  The slope of the
color-magnitude variations of Mira stars suggest that these stars,
along with the fainest, reddest LPV (V13) could be cluster Mira
variables.  Ultimately, it is difficult to assign definitive
membership designations since our observations do not provide an
accurate estimate of the stars' magnitudes and colors averaged over
their pulsation cycles.  Further monitoring of the LPVs in NGC~6316 is
required to improve the membership analysis.  Such monitoring would
also provide perriods and light curves for the LPVs that would enable
their classification as Irregular, Semi-Regular, or Mira variables.

The positions of the eclipsing binary candidates in Figure
\ref{fig_varcmd} suggest that these stars are associated with the
field population.  One exception may be V19, which lies only 1.5
arcmin from the cluster center.  In the CMD, this variable is in the
correct place to be an RRL member of the cluster.  However, its light
curve suggests it is a contact eclipsing binary star or perhaps a
Population II Cepheid (BL Herculis type).  If the latter, it should
have a luminosity $\sim$1 mag above the HB.  Its presence at the
apparent magnitude of the cluster HB would argue that it is located
well behind the cluster.  Whether V19 is an eclipsing binary or a BL
Her star, its membership in NGC~6316 seems unlikely.

The membership of the four probable RRL is of particular interest.
Only one of the stars, V17, lies within the cluster's tidal radius.
This star is about 0.5 mag brighter than the NGC~6316 red clump stars,
and is about the same brightness as the red clump of the Galactic
bulge (see \S \ref{sec_cmd}).  This argument favors membership in the
foreground bulge.  However, the RRL in the metal-rich globulars
NGC~6388 and NGC~6441 are unusually luminous \citep{pritzl02, lrww99,
pritzl01}, and their ``horizontal'' branches slope upward toward the
blue end \citep{rich97}.  Thus, it is possible that V17 is a luminous
member of NGC~6316.  The mean periods of the luminous RRL in NGC~6388
and NGC~6441 are unusually long ($\langle P_{RRab} \rangle \approx
0.75$ days) compared to those of bulge RRL ($\langle P_{RRab} \rangle
\approx 0.55$ days, \citet{alard96}).  Unfortunately, this test does
not yield a clear result, since V17 is well-fit by two periods: 0.4883
days and 0.9769 days.  The first period suggests a bulge origin, while
the second is consistent with cluster membership in the long-period,
metal-rich RRL paradigm.  We encourage additional monitoring of V17.

\section{Conclusions} \label{sec_disc}

We have obtained time-series $VI$ photometry of stars within a
projected radius of 8 arcmin from the metal-rich globular cluster
NGC~6313.  Our photometry is in agreement with that of \citet{arm88}
and \citet{davidge92}, but is $\sim$0.35 mag fainter and redder than
that of \citet{hr99}.  Though our photometry extends several
magnitudes below the cluster's red clump, we find no evidence for an
extended, blue horizontal branch of the type found in the metal-rich
globular clusters NGC~6388 and NGC~6441.  Comparison of our CMD with
isochrones \citep{bertelli94} indicates consistency with the
metallicity and reddening values listed by \citet{harris96}, provided
the cluster's distance modulus is $(m-M)_0 = 15.50$ mag.
 
\begin{figure}
\plotone{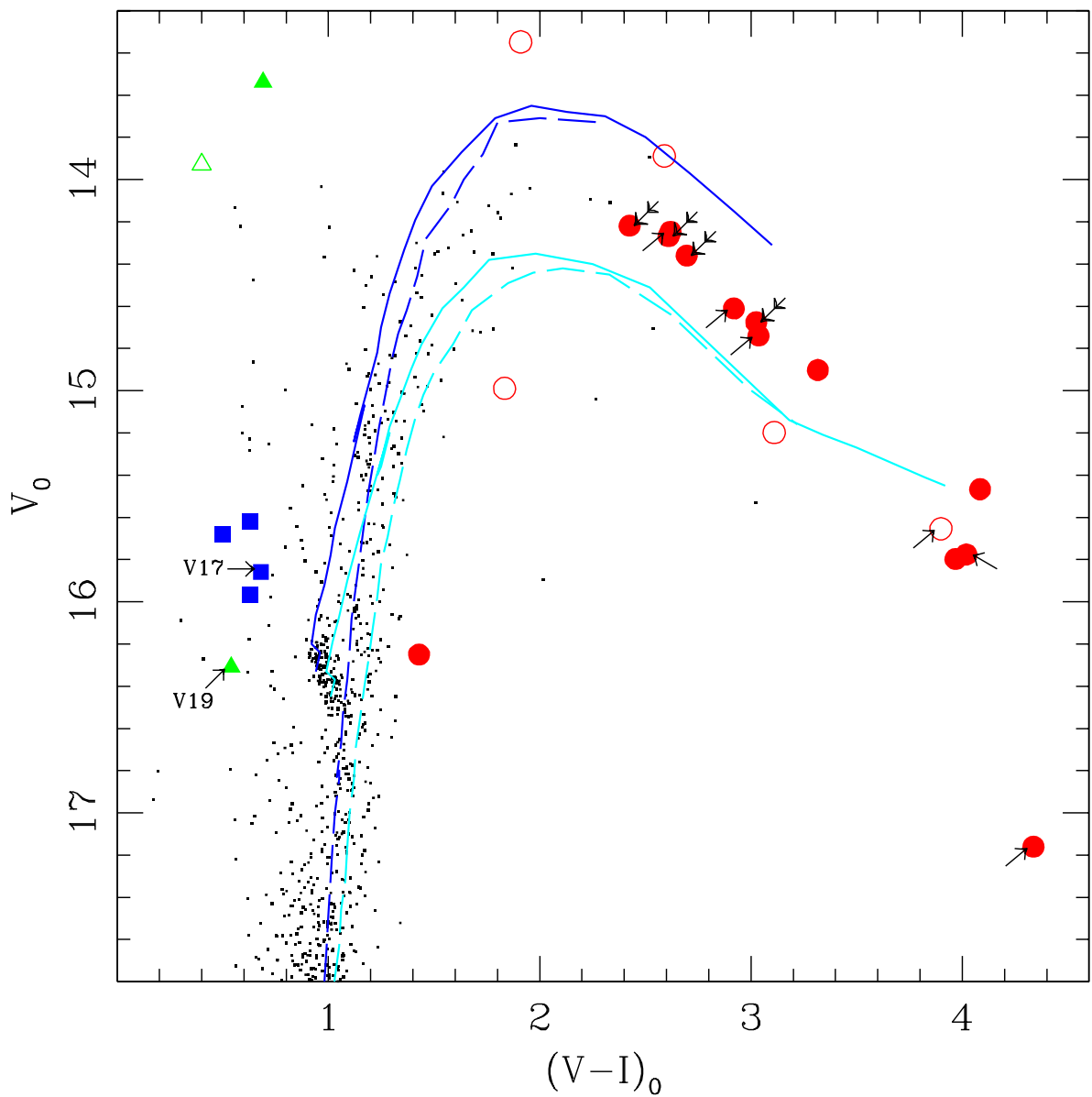}
\caption{The differentially dereddened CMD (see \S \ref{sec_redden})
is shown with the positions of the variable stars overlaid.  {\it
Circles} represent LPVs, {\it squares} indicate RRL, and {\it
triangles} mark eclipsing binary stars.  Filled and open symbols
represent probable and suspected variables, respectively.  The
suspected variable SV2 is marked with a cross.  Double and single
arrows point to variables within 0.71 and 3.0 arcmin of the cluster
center, respectively.  \citet{bertelli94} isochrones for [Fe/H] =
--0.7 ({\it upper set}) and --0.4 ({\it lower set}) are shown, shifted
by $(V-M_V)_0 = 15.50$ mag.  The isochrone RGBs are shown in {\it
dashed lines} and the AGBs are shown in {\it solid lines}.
\label{fig_varcmd}}
\end{figure}

We have detected over a dozen long period variable stars in the
direction of NGC~6316, and at least seven of them appear to be cluster
members.  Several of the member LPVs appear to lie above the RGB tip
in Figure \ref{fig_varcmd}, which would suggest they are Mira
variables according to the criteria of \citet{fw98}.  However, our
limited coverage of the stars' pulsation cycles leaves us uncertain
whether these LPVs are Irregular, Semi-Regular, or Mira type
pulsators.  If they are Mira stars, the specific frequency for Miras
in NGC~6316 should be comparable to that found in other globular
clusters of similar metallicity.  The specific frequency has been
defined as
\[S_x = N_x~10^{0.4(7.5 + M)},\]
where $N_x$ is the number of stars of type ``$x$'' in the cluster and
and $M$ is the cluster's integrated absolute magnitude.  Using the
Mira counts and $K$-band absolute magnitudes from Table 1 of
\citet{fw98}, we find that most clusters with --0.8 $<$ [Fe/H] $<$
--0.3 dex have $0.05 < S_{Mira} < 0.20$.  If we assume that $3 \pm 3$
of the very red LPVs in Figure 5 are Mira stars, then $S_{Mira} = 0.09
\pm 0.09$ for NGC~6316.  The agreement suggests that some or all of
these stars could be Mira members of this cluster.  Long-term
photometric monitoring is strongly encouraged to confirm this
interpretation.

We also detected four RR Lyrae variables along the line of sight
through the cluster.  Three lie outside the tidal radius of the
cluster, and their periods and magnitudes indicate they belong to the
foreground bulge.  Ambiguities in the pulsation period of the fourth
RRL (V17) leave us unable to determine whether it is a normal bulge
RRL projected on the cluster or a long-period, luminous member of
NGC~6316.  If V17 is a field star, the specific frequency of RRL in
NGC~6316 is clearly zero, whereas $S_{RR} = 0.35$ if V17 is the sole
RRL member of NGC~6316.  In this calculation, we used the integrated
$V$-band apparent magnitude and interstellar reddening values from
\citet{harris96} along with the distance modulus determined herein to
obtain $M_V = -8.65$ mag.  The $S_{RR}$ value for NGC~6316 is much
smaller than was found for the unusual clusters NGC~6388 and 6441,
which have $S_{RR} = 1.7$ \citep{pritzl02} and 7.5 \citep{pritzl01},
respectively.  By comparison, 47 Tuc has $S_{RR} = 0.3$
\citep{harris96}.  Whether or not V17 is found to be an RRL member of
NGC~6316, the process that produced abundant RRL and blue HB stars in
NGC~6388 and NGC~6441 is clearly not operating in NGC~6316.

\acknowledgments

The authors thank an anonymous referee for valuable comments and
suggestions.  This material is based upon work supported by the
National Science Foundation under Grant No. 9988259.  ACL also
acknowledges support by NASA through Hubble Fellowship grant
HF-01082.01-96A, which was awarded by the Space Telescope Science
Institute.  DLW and TMAW were supported, in part, by a Research Grant
from the Natural Engineering and Research Council of Canada (NSERC).





\clearpage

\begin{deluxetable}{rrrrrrrrrr}
\tabletypesize{\scriptsize} 
\tablecaption{Mean Photometry for Stars Near NGC~6316. \label{tab_clusphot}} 
\tablewidth{0pt} 
\tablehead{
\colhead{ID} & 
\colhead{$X_{pix}$} & 
\colhead{$Y_{pix}$} &
\colhead{$V$} & 
\colhead{$\epsilon_V$} & 
\colhead{$I$} &
\colhead{$\epsilon_I$} & 
\colhead{$N_V$} & 
\colhead{$N_I$} &
\colhead{$I_{WS}$} 
} 
\startdata 
22014 &  183.88 &  720.98 & 14.384 & 0.024 & 13.454 & 0.007 & 4 & 4 & 1.1 \\ 
64097 & 1275.35 & 1999.84 & 14.384 & 0.013 & 13.513 & 0.003 & 4 & 4 & 0.1 \\ 
3648 & 1427.78  &  128.13 & 14.418 & 0.005 & 13.222 & 0.005 & 4 & 4 & 0.5 \\ 
12836 & 1418.48 &  440.26 & 14.447 & 0.005 & 13.270 & 0.003 & 4 & 4 & -0.2 \\
60056 & 1293.93 & 1861.65 & 14.593 & 0.008 & 13.628 & 0.003 & 4 & 4 & -1.0 \\ 
\enddata



\tablecomments{The complete version of this table is in the electronic
edition of the Journal.  The printed edition contains only a sample.}

\end{deluxetable}

\clearpage

\begin{deluxetable}{rrrrrrr}
\tabletypesize{\scriptsize}
\tablecaption{Mean Photometry for Stars in Control Field. \label{tab_fieldphot}}
\tablewidth{0pt}
\tablehead{
\colhead{ID} 		& 
\colhead{$X_{pix}$}   	& 
\colhead{$Y_{pix}$}   	&
\colhead{$V$} 		&
\colhead{$\epsilon_V$}  	& 
\colhead{$I$} 		& 
\colhead{$\epsilon_I$} 
}
\startdata
51776 &  180.02 & 1580.01 & 14.773 & 0.015 & 13.736 & 0.013 \\
15006 &  663.02 &  478.49 & 14.784 & 0.012 & 13.679 & 0.011 \\
33217 &  197.75 & 1012.57 & 14.798 & 0.008 & 13.719 & 0.007 \\
34546 & 1244.87 & 1052.59 & 14.815 & 0.008 & 13.771 & 0.008 \\
 7412 & 1116.96 &  244.06 & 14.856 & 0.009 & 13.751 & 0.011 \\
 \enddata



\tablecomments{The complete version of this table is in the electronic
edition of the Journal.  The printed edition contains only a sample.}

\end{deluxetable}

\clearpage

\begin{deluxetable}{rrrrrrrrrl}
\tabletypesize{\scriptsize}
\tablecaption{Photometric Properties of LPV Stars. \label{tab_lpvphot}}
\tablewidth{0pt}
\tablehead{
\colhead{Name} 		&
\colhead{ID} 		& 
\colhead{$R$}   	& 
\colhead{$\overline{I}_M$}   	&
\colhead{$\overline{V-I}_M$}   	&
\colhead{$N_M$} 		&
\colhead{$\Delta I_{M-J}$}  	& 
\colhead{$\Delta (V-I)_{M-J}$} 		& 
\colhead{$N_J$}		&
\colhead{Comment\tablenotemark{a}} 
}
\startdata
V1  & 38583 & 103 & 12.49 & 3.20 & 3 & -0.05 & -0.09 &  7 & m  \\
V2  & 31501 &  30 & 12.51 & 3.17 & 3 & -0.06 & -0.14 &  6 & m  \\
V3  & 35496 &  42 & 12.52 & 3.21 & 3 & -0.05 & -0.24 &  7 & m  \\
V4  & 27562 &  50 & 12.72 & 3.64 & 4 &  0.29 &  0.19 &  5 & m  \\
V5  & 29770 &  21 & 12.65 & 3.83 & 3 &  0.23 &  0.35 &  6 & m   \\
V6  & 29391 &  60 & 12.41 & 3.46 & 3 & -0.35 & -0.41 &  8 & m, S \\
V7  & 56815 & 368 & 12.51 & 3.98 & 3 &  0.09 &  0.06 &  6 &    \\
V8  & 13811 & 256 & 12.37 & 4.76 & 3 &  0.20 &  0.09 &  6 &    \\
V9  & 32682 &  20 & 12.60 & 2.98 & 3 & -0.17 & -0.18 &  9 & m \\ 
V10 & 21530 & 121 & 12.57 & 4.58 & 3 & -0.15 & -0.14 &  8 &    \\
V11 & 32140 & 347 & 12.64 & 4.53 & 3 & -0.14 & -0.13 &  9 &    \\
V12 &  8488 & 348 & 15.79 & 2.06 & 7 &  0.18 &  0.01 & 12 & e, S \\
V13 & 43903 & 143 & 14.13 & 4.92 & 6 &  0.85 & -0.09 & 11 & S  \\
 \enddata


\tablenotetext{a}{Key: m = probable cluster member, S = significant
short-term variation, e = possible eclipsing binary.}


\end{deluxetable}

\clearpage

\begin{deluxetable}{rrrrrrrrl}
\tabletypesize{\scriptsize}
\tablecaption{Photometric Properties of Suspected Variable Stars. \label{tab_suspvar}}
\tablewidth{0pt}
\tablehead{
\colhead{Name} 		&
\colhead{ID} 		& 
\colhead{$R$}   	& 
\colhead{$\overline{V}$}   	&
\colhead{$\overline{V-I}$}   	&
\colhead{$\delta V$}  	& 
\colhead{$\delta I$} 		& 
\colhead{$N_{obs}$}		&
\colhead{Comment} 
}
\startdata
SV1  & 40992 & 136 & 17.14 & 4.53 & 0.19 & 0.07 & 11 & LPV? \\ 
SV2  & 16039 & 222 & 15.44 & 1.03 & 0.17 & 0.18 & 20 & short period \\
SV3  &  6438 & 444 & 16.39 & 2.50 & 2.60 & 3.15 & 18 & LPV? \\ 
SV4  & 15154 & 361 & 15.40 & 3.22 & 0.21 &  --  & 21 & LPV? \\ 
SV5  &  7526 & 409 & 16.71 & 3.74 & 0.38 &  --  & 21 & LPV? \\ 
SV6  & 12600 & 230 & 14.86 & 2.54 & 0.12 &  --  & 21 & LPV? \\ 
 \enddata




\end{deluxetable}

\clearpage

\begin{deluxetable}{rrrrrrrr}
\tabletypesize{\scriptsize}
\tablecaption{Time-Series Photometry of Variable Stars. \label{tab_timeser}}
\tablewidth{0pt}
\tablehead{
\colhead{Name} 		&
\colhead{HJD} 		& 
\colhead{$V$}   	& 
\colhead{$\epsilon_V$}   	&
\colhead{$I$}   	&
\colhead{$\epsilon_I$}  	& 
\colhead{$\phi$} 		& 
\colhead{$Q$}
}
\startdata
V1 & 2450225.6418 & 15.705 & 0.009 & 12.481 & 0.007 & 9.990 & 2 \\
V1 & 2450226.6308 & 15.690 & 0.009 & 12.505 & 0.011 & 9.990 & 2 \\
V1 & 2450227.7814 & 15.673 & 0.009 & 12.473 & 0.008 & 9.990 & 2 \\
V1 & 2450258.5347 & 15.858 & 0.009 & 12.544 & 0.005 & 9.990 & 4 \\
V1 & 2450258.5401 & 15.859 & 0.009 & 12.537 & 0.005 & 9.990 & 3 \\
V1 & 2450258.8264 & 15.867 & 0.005 & 12.553 & 0.007 & 9.990 & 2 \\
V1 & 2450259.4997 & 15.847 & 0.005 & 12.558 & 0.004 & 9.990 & 4 \\
V1 & 2450260.6096 & 15.844 & 0.005 & 12.531 & 0.008 & 9.990 & 4 \\
V1 & 2450262.5007 & 15.817 & 0.005 & 12.530 & 0.004 & 9.990 & 4 \\
V1 & 2450262.7478 & 15.804 & 0.006 & 12.520 & 0.006 & 9.990 & 2 \\
V1 & 2450263.4834 & 15.785 & 0.006 & 12.527 & 0.006 & 9.990 & 4 \\
V1 & 2450266.7368 & 15.747 & 0.007 & 12.509 & 0.007 & 9.990 & 2 \\
V2 & 2450225.6418 & 15.679 & 0.012 & 12.506 & 0.010 & 9.990 & 2 \\
V2 & 2450226.6308 & 15.677 & 0.014 & 12.524 & 0.011 & 9.990 & 2 \\
 \enddata



\tablecomments{The complete version of this table is in the electronic
edition of the Journal.  The printed edition contains only a sample.}

\end{deluxetable}

\clearpage

\begin{deluxetable}{rrrrrrrrrrrrl}
\tabletypesize{\scriptsize}
\tablecaption{Photometric Properties of Short Period Variable Stars. \label{tab_rrlphot}}
\tablewidth{0pt}
\tablehead{
\colhead{Name} 		&
\colhead{ID} 		& 
\colhead{$R$}   	& 
\colhead{$P$}   	&
\colhead{Class}		&
\colhead{$rms_V$} 	&
\colhead{$rms_I$} 	&
\colhead{$\langle V \rangle$}  	& 
\colhead{$\langle I \rangle$}  	& 
\colhead{$\Delta V$} 		& 
\colhead{$\Delta I$} 		& 
\colhead{$N_{obs}$}		&
\colhead{Comment} 
}
\startdata
V14 &  8333 & 327 & 0.3969 & WUMa & 0.015 & 0.024 & 15.05 & 13.73 & 0.24 & 0.21 & 20 & field \\
... &  ...  & ... & 0.1984 & RRc  & 0.020 & 0.028 & 15.08 & 13.76 &
0.29 & 0.21 & 20 &  short $P$ for RRc \\
V15 & 29329 & 454 & 0.5602 & RRab & 0.018 & 0.041 & 17.13 & 15.87 & 0.71 & 0.41 & 21 & field \\
V16 & 45168 & 366 & 0.2725 & RRc  & 0.014 & 0.042 & 17.19 & 16.06 & 0.47 & 0.29 & 22 & -- \\
V17 & 45182 & 158 & 0.4883 & RRab & 0.026 & 0.032 & 17.37 & 16.06 & 0.84 & 0.70 & 20 &  \\
... &  ...  & ... & 0.9769 & RRab & 0.030 & 0.029 & 17.59 & 16.23 & 0.98 & 0.64 & 20 &  \\
V18 & 76037 & 370 & 0.5822 & RRab & 0.017 & 0.021 & 17.48 & 16.22 & 1.06 & 0.60 & 20 & field \\
V19 & 40886 &  88 & 0.4225 & WUMa & 0.016 & 0.034 & 17.82 & 16.65 & 0.29 & 0.32 & 20 & field \\
... &  ...  & ... & 2.076  & Cep & 0.026 & 0.032 & 17.84 & 16.68 & 0.30 & 0.34 & 20 & -- \\
 \enddata




\end{deluxetable}


\end{document}